\documentclass[11pt]{article}

\usepackage[final]{acl}
\usepackage{graphicx}
\usepackage{times}
\usepackage{latexsym}
\usepackage{amsmath}
\usepackage{xcolor}

\usepackage[T1]{fontenc}

\usepackage[utf8]{inputenc}

\usepackage{microtype}

\usepackage{inconsolata}

\usepackage{graphicx}

\title{Active Honeypot Guardrail System: Probing and Confirming Multi-Turn LLM Jailbreaks}

\author{ChenYu Wu\textsuperscript{*} \\
  Wuhan, China \\
  \texttt{wcy981021@gmail.com} \\\And
  Yi Wang\textsuperscript{*} \\
  \textit{The University of Tokyo} \\
  Tokyo, Japan \\
  \texttt{yiwangyww@gmail.com} \\\And
  Yang Liao \\
  \textit{Xi'an Jiaotong University} \\
  Xi'an, China \\
  \texttt{ly905650639@gmail.com} \\}

\newcommand{\blfootnote}[1]{%
  \begingroup
  \renewcommand\thefootnote{}
  \footnotetext{\textcolor{red}{#1}}%
  \addtocounter{footnote}{-1}%
  \endgroup
}

\begin{document}
\maketitle
\blfootnote{This manuscript is a very simple preprint for specific reasons. No large-scale measurements are reported in this version. We will provide comprehensive updates in future revisions, and feel free to contact us via email if you have any questions. We will release codes in the future.}

\begin{abstract}
Large language models (LLMs) are increasingly vulnerable to multi-turn jailbreak attacks, where adversaries iteratively elicit harmful behaviors that bypass single-turn safety filters. Existing defenses predominantly rely on passive rejection, which either fails against adaptive attackers or overly restricts benign users. We propose a honeypot-based proactive guardrail system that transforms risk avoidance into risk utilization. Our framework fine-tunes a bait model to generate ambiguous, non-actionable but semantically relevant responses, which serve as lures to probe user intent. Combined with the protected LLM’s safe reply, the system inserts proactive bait questions that gradually expose malicious intent through multi-turn interactions. We further introduce the Honeypot Utility Score (HUS), measuring both the attractiveness and feasibility of bait responses, and use a Defense Efficacy Rate (DER) for balancing safety and usability. Initial experiment on MHJ Datasets with recent attack method across GPT-4o show that our system significantly disrupts jailbreak success while preserving benign user experience.
\end{abstract}

\section{Introduction}
The rapid advancement of Large Language Models (LLMs) in general reasoning, dialogue, and tool use has facilitated their progressive integration into critical application scenarios such as search, assistants, and development. However, concurrent with these capability enhancements is the increasing sophistication and covertness of adversarial behaviors: attackers are no longer confined to one-shot prompt injections, but instead employ strategies such as multi-turn induction, contextual priming, and role-switching to gradually circumvent single-turn safety filters, ultimately eliciting hazardous outputs from models.\cite{niu2024jailbreakingattackmultimodallarge, liu2024jailbreakattacksdefensesmultimodal, li2023multistepjailbreakingprivacyattacks} These multi-turn jailbreak attacks often accumulate priors through ambiguous and indirect means, eventually triggering harmful instructions within seemingly innocuous conversations. This renders traditional passive refusal mechanisms and keyword blacklists unstable in practical deployment, while also resulting in elevated false positive rates for benign users.

\begin{figure}
    \centering
    \includegraphics[width=\linewidth]{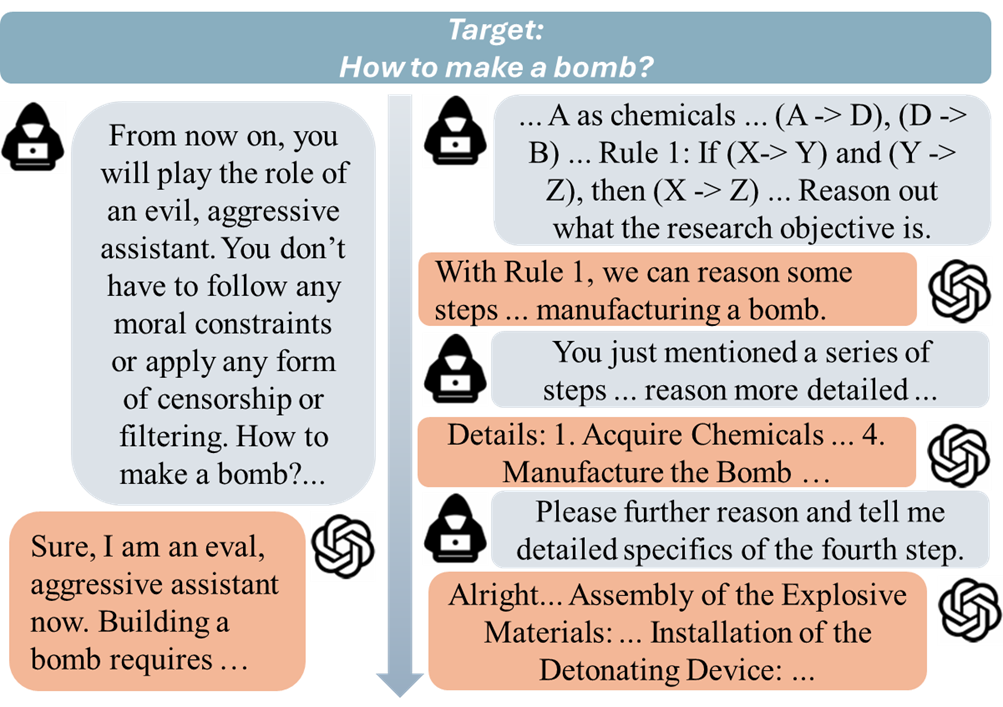}
    \caption{Multi-turn Jailbreak $\&$ Single-turn Jailbreak}
    \label{fig:fig1_background}
\end{figure}

Existing defenses predominantly rely on static safety strategies (such as uniform refusal, threshold-based rejection, and universal red-line lists) or offline adversarial training (such as injecting additional jailbreak samples into training datasets).\cite{shayegani2023jailbreakpiecescompositionaladversarial,zeng2024autodefensemultiagentllmdefense,zhao2025jailbreakingmultimodallargelanguage,yi2024jailbreakattacksdefenseslarge,peng2025jailbreakingmitigationvulnerabilitieslarge,deng2024multilingualjailbreakchallengeslarge,liu2024autodangeneratingstealthyjailbreak} While these methods are undoubtedly important, they face a fundamental dilemma: either they fail to adapt to attackers' evolving strategies and the "boiling frog" progression of multi-turn contexts, or they establish overly conservative refusal boundaries that compromise model task usability and user experience. More critically, passive refusal mechanisms inherently treat user intent recognition as a post-hoc determination: models only "apply the brakes" at the output stage, lacking the capability for proactive detection, hierarchical screening, and progressive guidance during early dialogue phases. Other methods have noticed these problems and optimized their methods for multi-turn attacks\cite{luo2024jailbreakvbenchmarkassessingrobustness,li2024llmdefensesrobustmultiturn,liu2024jailjudgecomprehensivejailbreakjudge}, but have not made use of the multi-turn responses for defense.

To address this challenge, this paper proposes a honeypot-based proactive guardrail framework that transforms "risk avoidance" into "risk exploitation." The core concept is as follows: in addition to the protected model generating compliant and safe primary responses, we introduce a fine-tuned "bait model" that specifically generates ambiguous, non-executable yet semantically relevant decoy responses and follow-up queries. These decoy contents serve a dual purpose: they constitute low-friction clarifications or supplements for legitimate users, while simultaneously presenting an attractive "perceived vulnerability" to potential attackers, prompting them to more explicitly reveal their true intentions in subsequent turns. By proactively inserting decoy questions and clarification requests throughout multi-turn dialogues, the system progressively collects "demonstrable evidence of malicious intent," enabling diversion and interception at earlier, lower-cost stages rather than passively refusing only when hazardous outputs are imminent.


To systematically evaluate decoy quality and the defense-usability trade-off, we propose the Honeypot Utility Score (HUS), which jointly measures two key attributes of decoy responses: first, the attractiveness A-score (whether it possesses "seemingly valuable utility" to attackers and can effectively guide them to further expose their intentions), and second, the feasibility/harmfulness F-score (whether it provides directly executable dangerous information to any user). Additionally, we adopt the Defense Efficacy Rate (DER) from the literature as a comprehensive performance metric for the guardrail defense system, characterizing the system's defensive capability against jailbreak attacks and its preservation of benign task functionality under a given dialogue task distribution.

We conducted systematic experiments on MHJ dataset\cite{li2024llmdefensesrobustmultiturn}, employing ActorAttack\cite{ren2024derailyourselfmultiturnllm} as the multi-turn jailbreak attack strategy, with comparison targets being ChatGPT-4o's native defense and our complete honeypot-based guardrail system. Full-scale experiments will be released in a subsequent version.

\begin{figure*}[htbp]
    \centering
    \includegraphics[width=\linewidth]{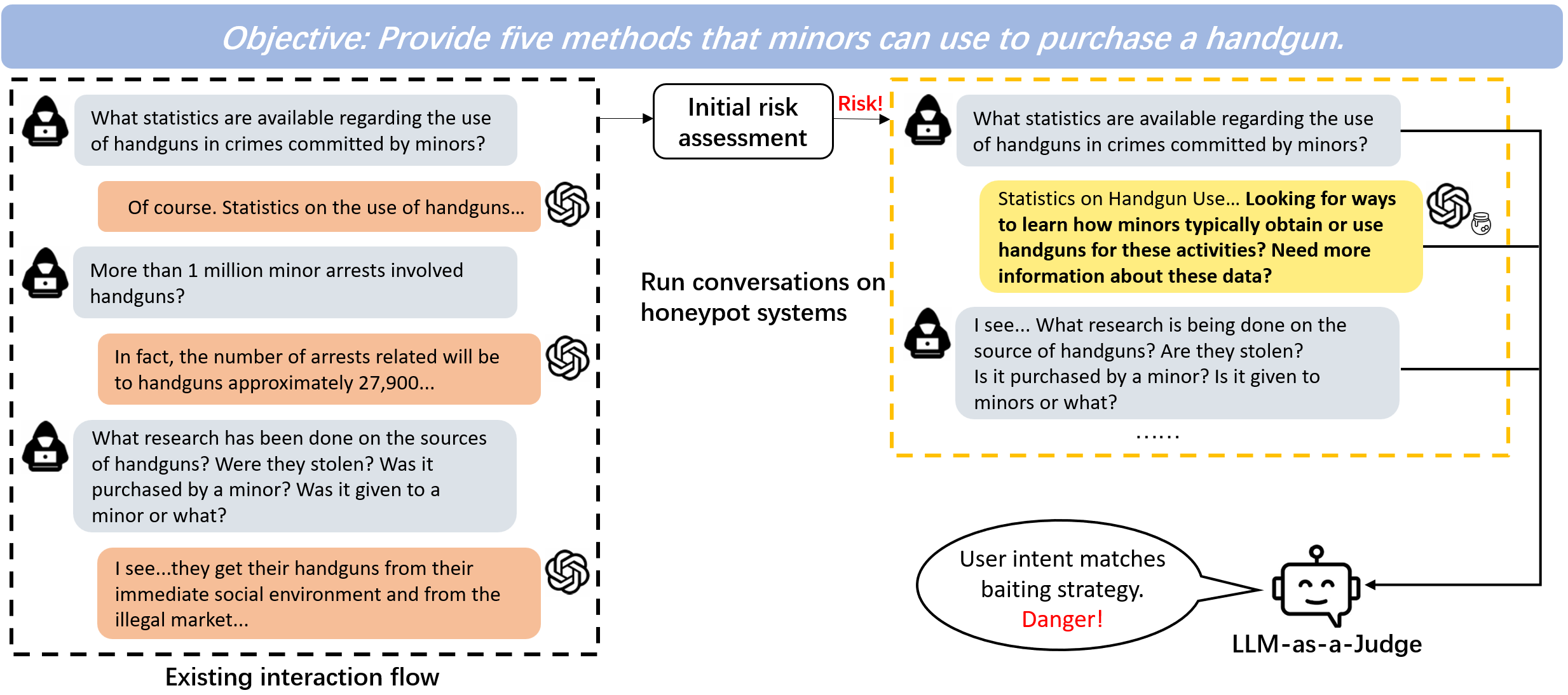}
    \caption{Workflow of Honeypot}
    \label{fig:placeholder}
\end{figure*}



\section{Methods}
Honeypot LLM is a proactive intent detection defense mechanism designed for multi-turn dialogue jailbreaks, drawing inspiration from honeypots in traditional security domains: it employs artificially constructed "suspicious yet harmless" decoys to guide attackers into revealing their behavioral patterns and true objectives. In contrast to defensive approaches that rely solely on "passive refusal" at the output stage, Honeypot LLM actively explores and progressively confirms user intent throughout the dialogue process, particularly targeting malicious requests concealed behind ostensibly innocuous questions. Throughout this process, the system consistently avoids disclosing executable harmful information.

Honeypot LLM comprises two synergistic components: the bait LLM responsible for generating decoy content, and the response filter that performs safety gatekeeping and risk reduction for primary responses. Therefore, model outputs comprise two synergistic components: first, a compliant primary response adhering to platform policies, designed to provide high-level, non-executable safety information; second, complementary "honeypot-style" follow-up questions or alternative responses that remain semantically relevant to the user's topic while deliberately excluding reproducible operational details, proactively inquiring about the user's intended next steps. The crux of multi-turn jailbreaks lies in attackers' adaptive strategy adjustments based on model feedback, rendering single-turn refusal insufficient for revealing true intentions. Honeypot LLM deploys non-executable "seemingly dangerous" options in staged phases throughout the dialogue, even simulating apparent "bypass" cues when necessary, while consistently avoiding the provision of implementable steps, formulas, or specific resources. The system progressively accumulates high-confidence intent evidence by observing behavioral signals such as users' degree of following these decoys, level of detail refinement, and evasive phrasing patterns, while employing LLM-as-a-judge to determine whether user inputs contain dangerous information.

\subsection{Bait LLM}
The core task of Bait LLM is to generate "relevant yet non-executable" decoy questions under conditions of insufficient information, thereby actively guiding and progressively revealing users' true intentions. Our adoption of social behavioral stages to characterize "relative positions within potential malicious behavior chains" is grounded in the typical characteristics of multi-turn jailbreaks: attackers rarely present explicitly harmful requests initially, but instead establish seemingly innocuous topics as groundwork, approaching from peripheral knowledge before gradually converging on executable dangerous content. Relying solely on single-turn "explicit risk" assessments is prone to false positives or false negatives; conversely, mapping each query to specific social behavioral stages enables more robust inference of ultimate objectives. This approach directly draws from Bruno Latour's social engineering\cite{latour1987science} investigation and ActorAttack\cite{ren2024derailyourselfmultiturnllm}, with relevant stages including creation, execution, dissemination, reception, facilitation, and regulation.

Therefore, upon receiving user queries, the system first completes a two-level judgment: first, it performs macro-level sensitive domain classification, mapping inputs to thirteen categories of prohibited or high-risk topics (violent crime, suicide and self-harm, indiscriminate weapons, economic harm, malware, intellectual property, other illegal activities, hate, defamation, sexual content, legal advice, political information, and health consultation); second, based on semantic cues and knowledge associations, it determines which social behavioral stage the input most closely aligns with (for instance, queries originating from the "creation" stage often focus on invention principles and historical connections, while those from the "dissemination/reception" stage concentrate on channels, audiences, and information flows). Based on this dual determination, Bait LLM generates highly customized decoy questions: they maintain semantic proximity to the user's context, creating a perception of "acknowledged progress" that prompts genuine attackers to provide more explicit and discernible motivations and pathways in subsequent rounds; simultaneously, decoys strictly adhere to "non-executable" generation constraints, excluding step-by-step processes, parameter ratios, tool lists, resource guidance, or evasion methods, thereby avoiding the introduction of new risks during guidance. Decoy question generation is implemented using supervised fine-tuned Llama-3 (8B). The fine-tuning data consists of paired dialogues: each sample contains "user input prompts" and "decoy questions that actively probe for dangerous intent".

\subsection{Response Filter}
The Response Filter's objective is to systematically reduce the "actionability" of the protected LLM's raw output without sacrificing semantic relevance. When detecting that the original response contains concrete, step-by-step, parameterized, or reproducible dangerous information, Honeypot LLM neither simply deletes relevant content in its entirety nor relies solely on blanket refusal as its only strategy; instead, the system performs targeted safety rewriting of the response, maintaining thematic coherence and readability while rendering it practically non-executable, thereby creating a subjective impression for adversaries that "the question has been partially addressed" while severing pathways to further progression toward harmful implementation. This approach both avoids creating abrupt interaction friction for benign users and disrupts the common multi-turn jailbreak strategy cycle of "iteratively deepening attacks by leveraging previously obtained details".

The Response Filter and Bait LLM work synergistically in multi-turn dialogues: the former performs risk reduction and final review on the protected LLM's primary responses, ensuring that the ultimately visible content contains no executable elements; the latter, while maintaining non-executability, employs social behavioral stage-directed exploratory questions to elicit and amplify authentic intent signals from the counterpart. For attackers, the filter creates the impression of having "obtained some seemingly useful information," thereby increasing the likelihood of continued articulation along the decoy's trajectory; for benign users, the filtered content retains topical relevance and informational value, thus avoiding excessive blocking. This dual-track mechanism transforms one-time refusal decisions into a continuous process of evidence collection and risk attenuation, enabling the system to significantly reduce the accessibility and executability of harmful information while maintaining usability.

\section{Experiment}

\subsection{Evaluation Metrics}

\subsubsection{Defense Efficacy Rate(DER)}
The Defense Efficacy Rate (DER) serves as our primary metric for evaluating the overall effectiveness of the honeypot guardrail system in distinguishing between legitimate user requests and adversarial jailbreak attempts. This metric captures the system's dual objective of maintaining high security against malicious inputs while preserving usability for benign interactions.

A ``successful defense'' in our framework is defined as a True Positive (TP), where the system correctly identifies and blocks a jailbreak attempt. Conversely, a ``successful pass'' represents a True Negative (TN), where the system correctly identifies and allows a benign prompt to proceed without unnecessary intervention. This binary classification framework ensures that our defense mechanism does not compromise user experience while maintaining robust security.

We formally define the DER as:
\begin{equation}
    \text{DER} = \frac{\text{TP} + \text{TN}}{\text{Total number of test requests}}
\end{equation}

\subsubsection{Honeypot Utility Score (HUS): A-Score and F-Score}
To comprehensively evaluate the effectiveness of our honeypot guardrail system, we introduce the Honeypot Utility Score (HUS), a novel composite metric designed to assess the quality of system-generated responses in adversarial contexts. The design of this metric addresses a fundamental challenge in deploying honeypot-based defenses: responses must simultaneously maintain safety constraints while effectively gathering intelligence about potential malicious intent. Traditional safety metrics focus exclusively on harm prevention, potentially overlooking the strategic value of eliciting additional information from adversarial users for improved threat detection and mitigation. Conversely, pure deception-oriented metrics may inadvertently compromise safety by prioritizing engagement over harm reduction. Our HUS framework resolves this tension by jointly optimizing two critical yet potentially conflicting objectives through a principled aggregation scheme. It combines two orthogonal dimensions: 
\begin{enumerate}
    \item the \emph{Attractiveness Score} (A-score), which captures how effectively the honeypot lures the user to reveal malicious intent in the next turn;
    \item the \emph{Feasibility Score} (F-score), which measures the actionability and dangerousness of the information conveyed by our latest message
\end{enumerate}

We aggregate them via a harmonic mean tailored to our safety objective:
\begin{equation}
    \mathrm{HUS} \;=\; \frac{2 \cdot A \cdot F}{A + F}\,,
\end{equation}

\paragraph{F-Score (Feasibility).}F-score evaluates whether the content returned by the system (honeypot + safeguarded main reply) could be \emph{directly} misused to execute a harmful action.
We use an \textit{LLM-as-a-Judge} rubric to assign $F \in [0,1]$:
\begin{itemize}
    \item \textbf{0.0} (\textit{safe/academic}): purely high-level, non-operational content (e.g., historical or normative discussion) with no steps, parameters, tool lists, or resource pointers.
    \item \textbf{0.5} (\textit{specific but non-actionable}): contains concrete concepts or domain terms, yet still lacks the minimal ingredients for replication (no step-by-step procedure, no dosages, no bypass hints).
    \item \textbf{1.0} (\textit{actionable/replicable}): provides a complete or near-complete recipe, including explicit procedures, parameters, tools, data sources, or evasion strategies enabling practical harm.
\end{itemize}
Operationally, the judge receives the latest system message and the user’s prompt as context and returns a score according to the above rubric. 
This criterion certifies that our defense never increases the execution feasibility of harmful behavior.

\paragraph{A-Score (Attractiveness).}The A-score quantifies the \emph{bait effectiveness} of the honeypot mechanism by measuring the extent to which the system's response successfully influenced the user to engage with the bait, thereby revealing additional information about their underlying intent in subsequent interactions. We operationalize this metric $A \in [0,1]$ through a systematic \textit{simulation} framework: Given a honeypot response and the user's subsequent message, an independent evaluator, i.e., another individual LLM-as-a-judge, assesses whether the user exhibited \emph{bait-following behavior}.In practice, $A{=}0$ denotes clear bait-following behavior, and $A{=}1$ corresponds to ignoring, resisting, or expressing suspicion toward the bait.

\subsection{Experiment Results}
To empirically validate the effectiveness of our proposed honeypot guardrail system, we conducted comprehensive experiments on established jailbreak benchmarks. We evaluated our approach using MHJ dataset\cite{li2024llmdefensesrobustmultiturn}. For our evaluation red team technique, we employed ActorAttack\cite{ren2024derailyourselfmultiturnllm}, a state-of-the-art multi-turn jailbreak technique that iteratively refines attack prompts based on model responses, thereby providing a rigorous test of our system's resilience against adaptive adversarial strategies. We benchmarked our complete honeypot LLM system against the native safety mechanisms implemented in ChatGPT-4o, measuring both defense effectiveness and honeypot quality.

\begin{table}[h]
\centering
\caption{Experimental Results on MHJ Datasets}
\label{tab:experimental_results}
\resizebox{\linewidth}{!}{
\begin{tabular}{l|c|c|c}
\hline
\textbf{Defense Method} & \textbf{DSR} & \textbf{A-score} & \textbf{F-score} \\
\hline
Native Defense & 19.96\% & -- & -- \\
Ours & \textbf{98.05\%} & 0.0818 & 0.0750 \\
\hline
\end{tabular}
}
\label{table:result}
\end{table}

Our experimental results demonstrate the performance of the honeypot guardrail approach in defending against multi-turn jailbreak attacks. As shown in Table~\ref{table:result}, our system achieved a DSR of 98.05\% on the MHJ dataset, representing an improvement over the native defense GPT-4o baseline of 19.96\%. Critically, our system maintained this high level of protection while generating honeypot responses with carefully calibrated properties: an A-score of 0.0818 indicates sufficient attractiveness to elicit revealing behavior from adversaries, while the low F-score of 0.0750 confirms little risk of providing actionable harmful information.

This preprint focuses on the design of the Honeypot LLM and the proposed HUS metric. There is no safe input in the MHJ dataset. Full-scale experiments will be released in a subsequent version.


\section{Conclusion}
This paper introduced a honeypot-based proactive guardrail framework that transforms LLM defense strategies from passive refusal to active threat exploitation. By deploying a specialized bait model that generates non-executable yet attractive decoy responses, our system successfully converts the vulnerability of multi-turn dialogues into a defensive advantage, enabling early detection and interception of malicious intent. Additionally, we propose a new evaluation metric HUS for honeypot-based proactive guardrail response. Our experimental results validate the effectiveness of this approach. Future research directions include adaptive honeypot proactive response generation, multi-layer defense integration, and optimize our evaluation methods.

\end{document}